# Investigation of the behavior of 5CB in a pore and a nano capsule

*M.A.Korshunov*

*L.V.Kirensky Institute of Physics Siberian Branch of the Russian Academy of Sciences, 660036 Krasnoyarsk, Russia*

e-mail: kors@iph.krasn.ru

We use a method of molecular dynamics to investigate the distribution of liquid crystal 5CB molecules in a polymeric matrix. Temperature dependences of parameters of an order <P2> are calculated. Calculations have shown that depending on the size of a capsule and on a time the transition temperature changes.

Liquid crystals consisting of a polymeric matrix dispersed by molecules of a liquid crystal, are perspective materials for practical use in areas of nonlinear optics, optoelectronics, the microwave technique and nano-photonics [1]. Advantage of such substances is management possibility their properties by change of superficial forces of interaction a liquid crystal-polymer. Is of interest to consider behaviour of molecules in capsule a liquid crystal for the different sizes of a capsule and in a time.

Calculations were done by a method of molecular dynamics [2].

Calculations have shown that if the sphere is homogeneous, molecules 5CB close to a surface (3-5 layers) settle down along borders enough упорядоченно. It is shown in Figure 1. With temperature increase most slowly disorder molecules located near borders and along a line connecting poles.

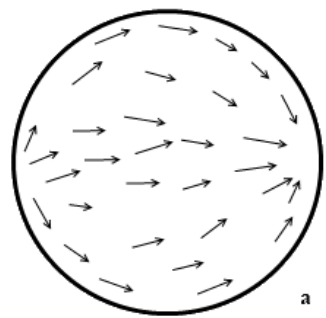

Figure 1. Arrangement of molecules in 5CB.

From the found arrangement of molecules at change of temperature for 5CB temperature dependence of parameter of an order <P2> (Figure 2) is calculated. Its behavior will qualitatively be co-ordinated with the data received of the experiment executed on porous structures [3]. In Figure 2 bottom curve corresponds to change of parameter of an order in a time in the size 15nm. The average curve in a capsule.

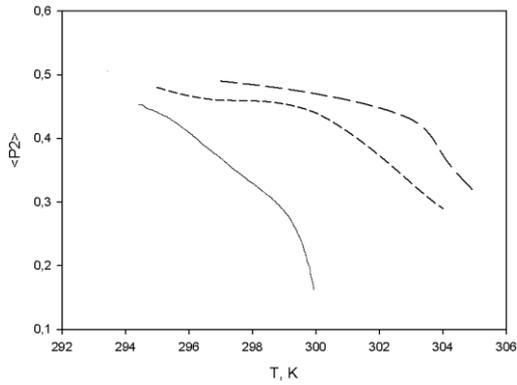

Drawing 2. Temperature dependence of parameter of an order <P2>.

Thus, the behavior of 5CB on a time and nano capsules is studied. Distribution of molecules of a liquid crystal paid off a method of molecular dynamics. Temperature dependences of parameter of an order <P2> are calculated. Calculations have shown that depending on the size of a capsule the transition temperature changes. At increase in its sizes the data on capsulated crystals comes nearer to the volume.

From graphs it is visible that in the capsulated liquid crystal the temperature of transition goes down a little.